\documentclass[sigconf]{acmart}

\def\BibTeX{{\rm B\kern-.05em{\sc i\kern-.025em b}\kern-.08emT\kern-.1667em\lower.7ex\hbox{E}\kern-.125emX}}
    
\acmConference[KDD '20]{KDD '20: Workshop on Knowledge Graphs and E-Commerce}{August, 2020}{San Diego, California}

\begin{document}

%
\title[Schema.org Best Practices]{On using Product-Specific Schema.org from Web Data Commons: An Empirical Set of Best Practices}

%

\author{Ravi Kiran Selvam}
\email{rselvam@isi.edu}
\affiliation{%
  \institution{USC Information Sciences Institute}
  \streetaddress{4676 Admiralty Way, Ste. 1001}
  \city{Marina del Rey}
  \state{California}
  \postcode{90292}
}

\author{Mayank Kejriwal}
\email{kejriwal@isi.edu}
\affiliation{%
  \institution{USC Information Sciences Institute}
  \streetaddress{4676 Admiralty Way, Ste. 1001}
  \city{Marina del Rey}
  \state{California}
  \postcode{90292}
}

%
\renewcommand{\shortauthors}{Anonymous}

%
\begin{abstract}
Schema.org has experienced high growth in recent years. Structured descriptions of products embedded in HTML pages are now not uncommon, especially on e-commerce websites. The Web Data Commons (WDC) project has extracted schema.org data at scale from webpages in the Common Crawl and made it available as an RDF `knowledge graph' at scale. The portion of this data that specifically describes products offers a golden opportunity for researchers and small companies to leverage it for analytics and downstream applications. Yet, because of the broad and expansive scope of  this data, it is not evident whether the data is usable in its raw form. In this paper, we do a detailed empirical study on the product-specific schema.org data made available by WDC. Rather than simple analysis, the goal of our study is to devise an empirically grounded set of best practices for using and consuming WDC product-specific schema.org data. Our studies reveal six best practices, each of which is justified by experimental data and analysis.
\end{abstract}

%
%


%

%

%
\maketitle

\section{Introduction}
Structured data has continued to play an increasingly important role for Web search and applications. In fact, according to Cafarella et al. \cite{cafarella2011}, the expanding quantity and heterogeneity of Web structured data has enabled new solutions to problems, especially concerning search engine optimization (SEO) and data integration spanning multiple web  sources. Two application  areas that have benefited greatly from structured data are e-commerce and advertisements. Because of structured data  and markup, it is now easier  than ever both to advertise and find products on the Web, in no small part due to the ability of search engines to make good use of this data. There is also limited evidence that structured data plays a key role in populating Web-scale knowledge graphs such as the Google Knowledge Graph that are essential to modern semantic search \cite{gkg}. 

However, even though most e-commerce platforms have their own proprietary datasets, some resources do exist for smaller companies, researchers and organizations. A particularly important source of data is schema.org markup  in webpages. Launched in the early 2010s by major search engines such as
Google and Bing, schema.org was designed to facilitate structured
(and even knowledge graph) search applications on the
Web. The Web Data Commons (WDC) project has crawled increasing
amounts of schema.org data in recent years \cite{wdc}, including in the e-commerce and products domain. WDC schema.org data has broad coverage at the level of `pay-level  domains' (such as rakuten.com), languages and product categories, providing a golden opportunity for researchers to use this data in downstream applications and analyses. 

Yet, there are also considerable challenges in using this data. Not including noise due to variations and misspellings, language tags on text literals  may be inaccurate, and there may be skew both in the distribution of languages and pay-level domains. Some of the properties may have less semantic validity than others. All of this is further made difficult by the fact that WDC product-specific schema.org datasets are non-trivial in size (with each year's data comprising hundreds of gigabytes even in compressed format), which precludes significant manual processing and labeling (or `eyeballing'). What is needed and is currently lacking is a set of empirically grounded best practices for consuming product-specific schema.org data released by the WDC. These best practices, once determined, could then be implemented in practice by any engineer looking to consume these datasets in applications and processes of their own. 

In this paper, we conduct a series of empirical studies to describe and justify such a set of practices. We use 2018 product-specific schema.org WDC for this purpose. In total, we conduct four broad studies investigating issues ranging from variation in schema.org concept representations to language inconsistencies and data localization. From our experiments, we distill six best practices for both researchers and practitioners. We show that, while extremely promising, schema.org data from WDC should be used while keeping these  best practices in mind to avoid issues of quality, scale and bias.  Future work may further refine these practices and supplement them significantly. 

The rest of this work is structured as follows. In Section \ref{relatedwork}, we provide some synthesis of related research. Section \ref{rawdata} provides details of  the raw data used for the study, while Section \ref{study} describes the methodology and results from each of our empirical studies. Section \ref{bestpractices} provides a summary of best practices distilled from these studies, and Section \ref{conclusion} concludes the paper.

\section{Related Work}\label{relatedwork}
The research presented in this paper is related to several of the existing lines of work that we briefly describe below. There has been a considerable amount of work on schema.org already, both in describing its principles and its evolution. For example, the authors in \cite{analyzingschemaorg} describe the core principles behind a plausible version of schema.org and state the formal semantics of using schema.org. In a related, but different vein, the authors in \cite{schemaorgevolution} perform large scale analysis of the usage of schema.org vocabularies over time.

A more closely related work is \cite{heuristics2015}, which describes the set of simple heuristics that could be applied to WDC microdata \cite{wdcdataset} so that consumers can use them to fix common mistakes as a post-processing step. The authors of \cite{hotelcorpususage} demonstrate a similar analysis of the validity of schema.org concepts in the hotel domain.

Good example of work that is more e-commerce oriented is \cite{productdataanalysis}, which describes the task of integrating the descriptions of electronic products from websites that use microdata markup to represent information and the various challenges that the authors faced. Yet another example is \cite{productmatchingdownstream}, which uses the structured data from the web as supervision for training feature extraction models to extract attribute-value pairs from textual descriptions of products.

Other examples of work that are related to schema.org analysis but that are too extensive to describe here include \cite{orgsemanticmarkup}, \cite{ontologyevolutionanalysis}, \cite{linkeddataanalysis}, \cite{lodlaundromat}, \cite{profilingrdfdata}, \cite{reconcilingwebdata} and \cite{schemaorgprofiling}.

\section{Raw Data}\label{rawdata}
As mentioned in the introduction, \emph{schema.org} \cite{schemaorg} is a collaborative effort by major search engines such Google, Yahoo, Microsoft, Yandex and open community members to create, maintain and promote schemas for  publishing embedded structured data on webpages. Schema.org has vocabularies that support different encoding schemes like microdata, RDFa and JSON-LD. Schema.org vocabularies are used by more than 10 million websites to add markup to their webpages. Schema.org markup helps the search engines better understand the information present in webpages much better, which in turn facilitates richer search experiences for search engine users.

The full extent of schema.org on the Web may not be known to any individual or organization beyond a large search engine such as Google. The Common Crawl is an initiative to allow researchers and the general public to have access to reasonably high-quality crawl data that was previously only available to major search engines. The schema.org portion of the Web Data Commons (WDC) project supports researchers and companies in exploiting the structured information available on the Web by extracting schema.org and other kinds of structured data from the webpages in the Common Crawl and making the data available. Conveniently, WDC also provides \emph{class-specific} subsets of the extraction corpus for a selection of schema.org classes. Such subsets only (or mostly) contain instances of a specific class (e.g., Products,  Books,  Movies etc.) which is especially convenient for domain-specific analysis.

We used the \emph{Product-specific} subset of the schema.org data contained in the November 2018 version\footnote{http://webdatacommons.org/structureddata/2018-12/stats/stats.html} of the Web Data Commons Microdata dataset. We will refer to this dataset as the \emph{Product dataset} in subsequent sections. The size of this dataset is roughly 112.7 GB in compressed form. The dataset may be downloaded as chunks with each compressed chunk being of size 1.2 GB. This dataset contains around 4.8 billion quads, 7.4 million URLs and 92,000 hosts. The top classes which are present in this dataset are shown in the Table \ref{tab:topClasses}.

\begin{table}[h!]
  \caption{Breakdown of the nodes by Entity class (only the 5 most frequent entity classes are shown) }
  \label{tab:topClasses}
  
  \resizebox{\columnwidth}{!}{%
   \begin{tabular}{|p{2.0in}|p{1.0in}|}
    \toprule
    Entity Class & Entity Count  \\
    \midrule
    http://schema.org/Product & 307.3 M \\
    \hline
    http://schema.org/Offer & 236.3 M \\
    \hline
    http://schema.org/ListItem & 65.7 M \\
    \hline
    http://data-vocabulary.org/Breadcrumb & 45.5 M \\
    \hline
    http://schema.org/AggregateRating & 30.4 M \\
  \bottomrule
   \end{tabular}}%

\end{table}

\section{Empirical Study}\label{study}
This section demonstrates the experiments conducted to study the Product dataset for determining the best practices to utilize the schema.org dataset for downstream tasks. 

\subsection{Variations of Schema Use}\label{sec:schemavar}
Website publishers, including organizations, use schema.org as a way to add structured markup to their websites. Given a schema.org concept such as ``https ://schema.org/product/name", publishers tend to use (whether inadvertently or not) variations of the concept such as the domain prefix ``bib.schema.org" depending on the information being annotated with markup, or misspellings of ``schema.org" such as ``schema.ofg". To utilize the Product dataset for downstream tasks, these variations make it difficult for the researchers who may need to extract information from triples associated with a particular concept mentioned in the standard schema.org or data-vocabulary.org vocabularies commonly used for publishing markup.  Our first, therefore, attempts to study these variations in more detail both to understand their nature and extent.

We design and conduct such a study by identifying 137 unique concepts in this dataset. The important problem over here is to differentiate between concepts and different variations of the same concept. We identified certain concepts as \emph{base concepts} based on correct spelling and common usage. We manually identified clusters of variations by creating a cluster for each base concept and manually assigning variations and misspellings of that base concept to belong to that cluster\footnote{We also validated our approach by using the DBSCAN clustering algorithm based on the Normalized Levenshtein Distance metric}. We identified two major clusters that represent the variations and misspellings of two concepts: schema.org and data-vocabulary.org. All other concepts are placed in a third cluster.

We conducted further analysis on each of the two major clusters to identify if there were \emph{consistent sources} of variation. We found that variations can be categorized along three dimensions, namely (i) \emph{variations in sub-domain} (e.g., \url{health-lifesci.schema.org} vs. \url{bib.schema.org} or \url{www.data-vocabulary.org} vs. \url{rdf.data-vocabulary.org}), (ii) \emph{misspellings in the second-level domain} (e.g., \url{ruschema.org} vs. \url{scheme.org} or \url{datavocabulary.org} vs. \url{data-vocabulary.org}), and (iii) \emph{misspellings in the top-level domain} (e.g., \url{schema.org.cn} vs. \url{schema.ofg}).

\begin{table}
  \caption{Count of product nodes for each of the sources of variation identified in two major clusters (schema.org and data-vocabulary.org)}
  \label{tab:freq}
   \resizebox{\columnwidth}{!}{
   \begin{tabular}{|p{1.7in}|p{0.7in}|p{0.7in}|}
    \toprule
    Source of variation & schema.org cluster & data-vocabulary.org cluster \\
    \midrule
    Base concept & 836 M & 49.3 M \\
    \hline
    Variations in sub domain & 4.5 M  & 1.5 M \\
    \hline
    Misspelling in Second-level domain & 8.6 K &   3.7 K\\
    \hline
    Misspelling in Top-level domain & 5.9 K & NIL \\
  \bottomrule
   \end{tabular}}
\end{table}

Compared to the number of nodes that belong to the two major clusters, the number of nodes having concepts from the third cluster is negligible and is not explored further. The variations that belong to the third cluster may be attributed to the errors in extraction or casual misuse of nomenclature by the users while adding markup to their websites. 

\subsection{Characterization of Product Data}\label{sec:productchar}
The websites that embed the information using schema.org markup are used for extraction to create the Product dataset. We hypothesize that not all the information that is present in the Product dataset is \emph{semantically valid}. For example, the product name property which is extracted for a particular product node might not be valid if it contains "Null" or "N/A" or if it contains another piece of information such as the URL (rather than the literal representing the name of the product).

\begin{figure*}
  \centering
  \includegraphics[width=\linewidth, height=8cm]{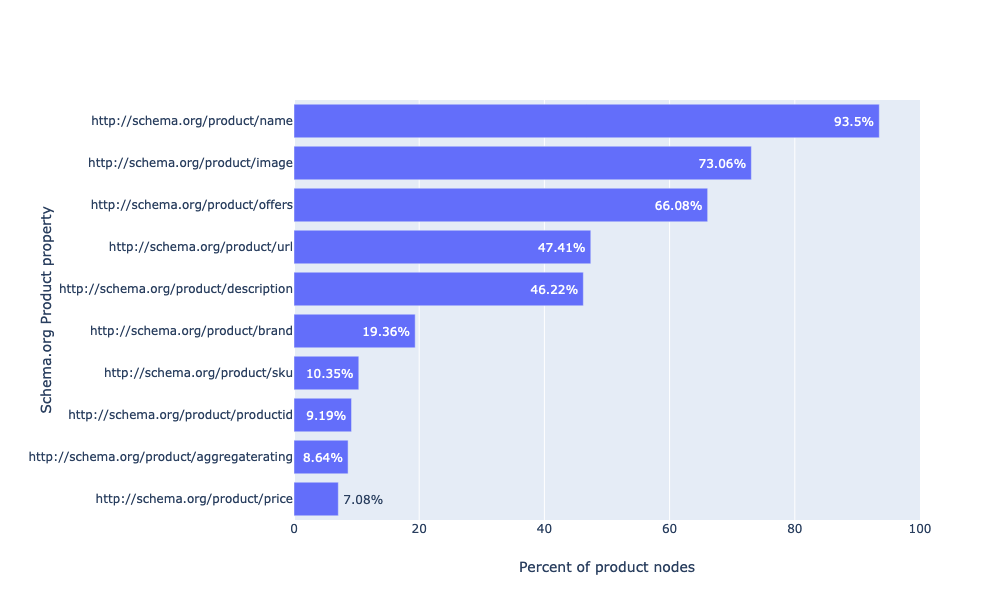}
  \caption{Percent of product nodes associated with a product property (only the 10 most frequent product properties are shown) }
  \label{fig:topproperties}
  \Description{Percent of product nodes associated with a product property (only the 10 most frequent product properties are shown) }
\end{figure*}

In this section, we study the semantic validity of properties of product nodes based on heuristic constraints. By heuristic constraints, we mean conditions that we intuitively associate with both the type of the property value or \emph{object} of the product property (e.g., text literal for some property values, vs. node or even URL as object). We determined the heuristic rules to check validity of the product properties by randomly sampling a subset of triples associated with each product property and examining them.  Since the number of unique product properties is very large, it is not practical to examine all the product properties; hence we limited our exploration to the ten most frequently occurring product properties. The heuristic rules that we devised to check the validity of the ten most frequently occurring product properties are described in Table \ref{tab:prodChar} and Figure \ref{fig:topproperties} describes the percent of product nodes associated with each of the 10 most frequent product properties.

While exploring the sample triples associated with product properties, we observed another kind of invalidity for certain properties like ``product name", ``brand" etc. Specifically, we found that even when the product property satisfies the heuristic rules previously described, the object still contains information that is associated with another property instead of the stated property. For example, we observed certain triples which contained a product ID or product description as objects  of the ``product name" property. These kinds of semantic errors are much harder to detect, and are left for future study. 

\begin{figure*}
  \centering
  \includegraphics[width=\linewidth, height=7cm]{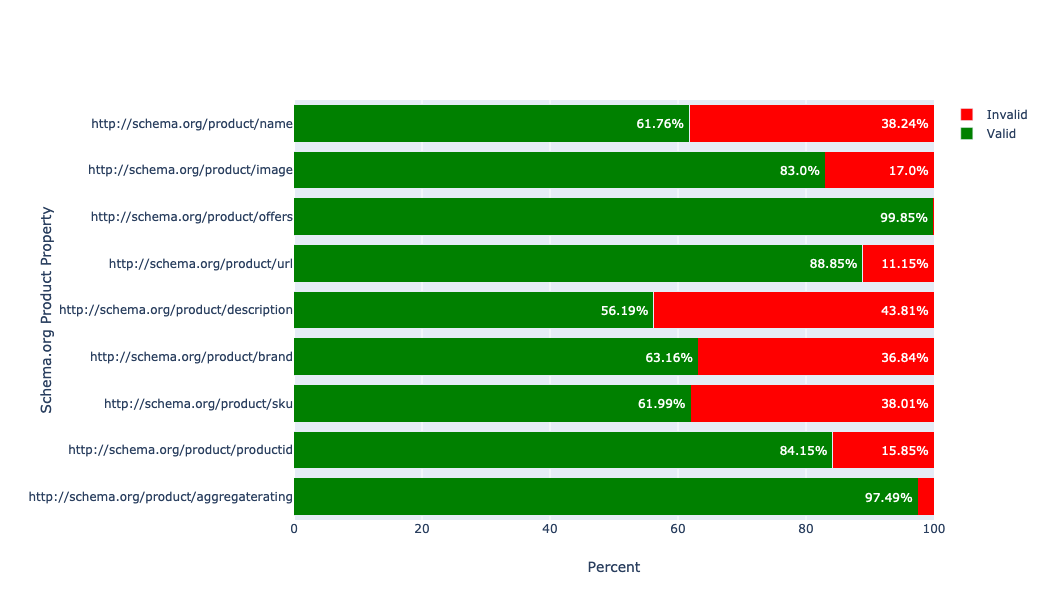}
  \caption{Breakdown of Valid vs. Invalid shares for each product property. Valid share of a product property is defined as the percent of product nodes with a property value that satisfies the heuristic rules of validity stated for that property. (only the 10 most frequent product properties except the "product price" property are shown)}
  \label{fig:propertyvalidity}
  \Description{Percent of product nodes associated with a product property (only the 10 most frequent product properties are shown) }
\end{figure*}

Furthermore, we define a product \emph{node} to be valid if it contains at least five valid properties out of the ten most frequently occurring properties and satisfies the heuristic rule which states that \emph{the length of the preprocessed text literal associated with product name property is less than the length of the preprocessed text literal associated with product description property}. Note that, in preprocessing text literals, we take standard steps such as removing extraneous white spaces such as tabs, new lines etc. We found over 32 million valid product nodes (which account to 10.66\% of the total product nodes in the Product dataset) that satisfies the heuristic rules for being a valid product nodes. Figure \ref{fig:propertyvalidity} describes the percent of product nodes being associated with property values that are determined to be valid by satisfying the heuristic rules for being valid properties as mentioned in Table \ref{tab:prodChar}.

\begin{table*}
\caption{Heuristic rules to check the semantic validity of properties associated with product nodes (only the 10 most frequent product properties except the "product price" property are considered; The "product price" property contains different currency symbols which are represented in unicodes and are challenging to validate) }
\begin{center}
\begin{tabular}{|p{1.0in}|p{5.5in}|}
    \toprule
    Product property & Heuristic rules for checking validity \\
    \midrule
    name, description & 
        Must be a text literal and have a non-zero length after removing white spaces. \newline
        Must not be an url. \newline
        Must not contain ("Null", "N/A"). 
    \\
    \hline
    image, url & 
        Must be a valid url. \newline
        Must not contain ("Null", "N/A"). 
    \\
    \hline
    offers &
        Must be a node (or) valid url. \newline
        Must not contain (text literals, "Null", "N/A"). 
    \\
    \hline
    brand & 
        Must be a non-zero length text literal (or) valid url (or) node. \newline
        Must not contain ("Null", "N/A"). 
    \\
    \hline
    sku, productid & 
        Must be a text literal and have a non-zero length after removing white spaces. \newline
        Must not contain ("Null", "N/A"). 
    \\
    \hline
    aggregaterating & 
        Must be a node. \newline
        Must not contain (text literals, "Null", "N/A"). 
    \\
  \bottomrule
\end{tabular}
\end{center}\label{tab:prodChar}
\end{table*}

\subsection{Data Skewness}\label{sec:dataskew}
Not all the languages and pay-level domains are equally represented in the Product dataset. It may be critical for certain NLP and multi-lingual applications to understand the distribution of languages and pay-level domains in the dataset. The dataset can be \emph{skewed} such that it has a significant portion of triples with text literals in a particular language, without also having a reasonable proportion of triples with text literals in other languages. This skew in language distribution might bias the results of downstream tasks that expect data in multiple languages with similar representation as in the real world (or the broader Web). We note that, like the rest of the experiments in  this paper, such a skew does not necessarily mean that the entire schema.org component of the Web is skewed; only  that the WDC Product dataset that we are studying is skewed. This is important to remember  despite the importance of WDC in any large-scale studies involving this kind of Web data. 

Furthermore, skew may arise not just at the language level but also at the level of pay-level domains associated with particular types of products (e.g., the pay-level domain \emph{fineartamerica.com} lists only products related to art like paintings and home decor  as opposed to clothing or electronics). This skew in pay-level domain distribution may bias the results of downstream tasks which need product data to be equally distributed among different categories. 

\begin{figure*}
  \centering
  \includegraphics[width=15cm, height=7cm]{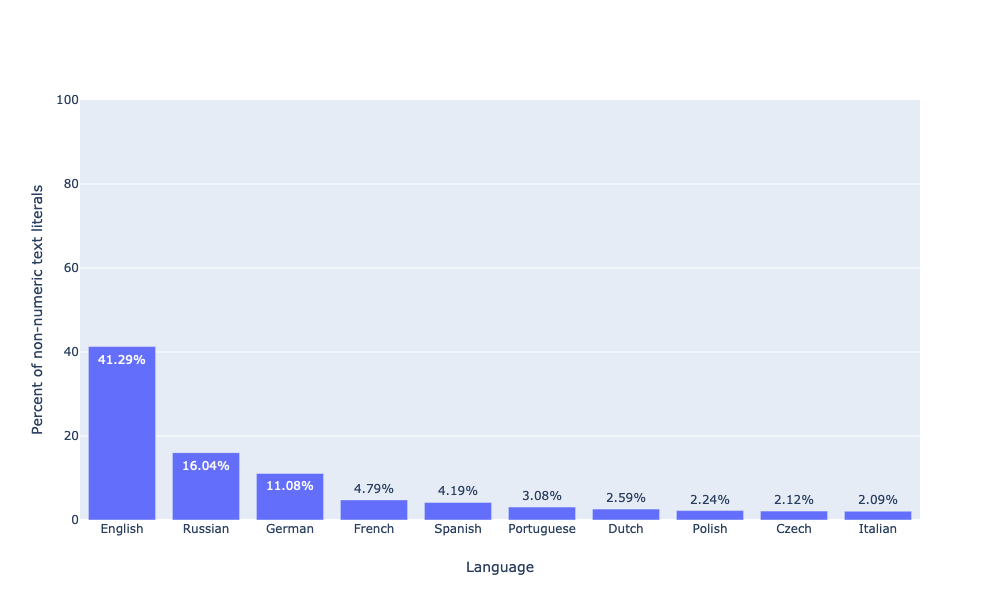}
  \caption{Breakdown of the non-numeric text literals by associated language tags (only the 10 most frequent languages are shown). A non-numeric text literal is one where at least 50\% of the characters as non-digits.}
  \label{fig:topLangs}
  \Description{Breakdown of the non-numeric text literals by associated language tags (only the 10 most frequent languages are shown). A numeric text literal is one where at least 50\% of the characters as non-digits.}
\end{figure*}

To study issues of skewness in  the Product dataset by first parsing and extracting the language tags that are explicitly associated with text literals by virtue of being represented in  RDF. We found 1,072 unique language tags in the dataset. We observed that many language tags are simply variations of the same language but associated with different countries. We reconciled all those language tags which represented the same language into a single cluster, yielding 249 unique language `clusters'. From Figure \ref{fig:topLangs}, we observe that the dataset is skewed towards having higher numbers of triples with text literals in English, followed by Russian and German.  

We investigated further to check if the language tags associated with text literals are consistent with the \emph{actual} language of the text literals, since the explicit tags could be incorrect. We randomly sampled a small subset of triples containing text literals and examined them manually. We found some evidence of disagreement between the actual language of the text literals and their associated language tags. Since it is not possible to manually peruse the whole dataset, we opted to design an alternate experiment to test inaccuracy in language tag declarations. 

Specifically, we used a pre-trained fastText-based language identifier model\footnote{The pretrained model can be downloaded from  https://fasttext.cc/docs/en/language-identification.html} which can recognize 176 languages in a fast and accurate manner. The language identifier was trained on data from Wikipedia, Tatoeba, and SETimes and achieved more than 93 \% accuracy on many standard language identification datasets from Wikipedia, TCL and EuroGov. Since the dataset is still too large for all text literals to be tagged using fastText-based language identifier without expending considerable computational resources, we randomly sampled 1/100th of the total number of text literals in each chunk of the dataset, yielding 100,000 samples per chunk. With 97 chunks, the total number of samples considered this evaluation is 10 million, which is large enough to  draw reasonable conclusions. We also found many text literals had numeric data like prices, dates, phone numbers, dimensions, model numbers, and time intervals. These numeric text literals are not associated with any particular language and will cause obvious problems in estimating the level of agreement between explicitly declared language tags and the outputs of the automatic language identifier. Hence, we removed these numeric text literals (text literals having at least 50\% of the characters as digits) from our evaluation by conducting some preprocessing. 

\begin{figure*}
  \centering
  \includegraphics[width=15cm, height=7cm]{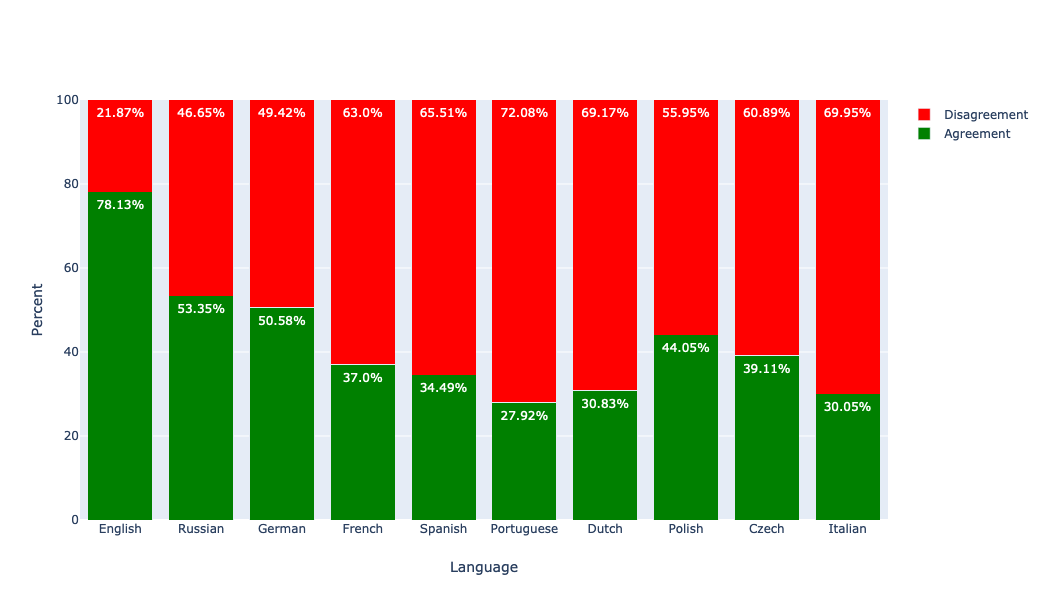}
  \caption{Breakdown of Agreement vs. Disagreement shares for each language. Agreement share of a language is defined as the percent of non-numeric text literals with an associated language tag that is consistent with the actual language of the text literals. }
  \label{fig:langagree}
  \Description{Breakdown of Agreement vs. Disagreement share for each language. Agreement share of a language is defined as the percent of non-numeric text literals with an associated language tag that is consistent with the actual language of the text literals. }
\end{figure*}

Figure \ref{fig:langagree} shows the levels of agreement and disagreement for the 10 most popular languages found in the dataset. We find that disagreement is much higher for some languages than others, but all languages have a non-trivial share of disagreement. In other words, for any quality-critical application, explicit language tags should not be directly trusted. Since the automatic identifier itself is not perfect, we recommend using only that subset of triples (or triples with text literals) where there is agreement between the automatically identified, and explicitly declared, language tags.

\begin{figure*}
  \centering
  \includegraphics[width=15cm, height=7cm]{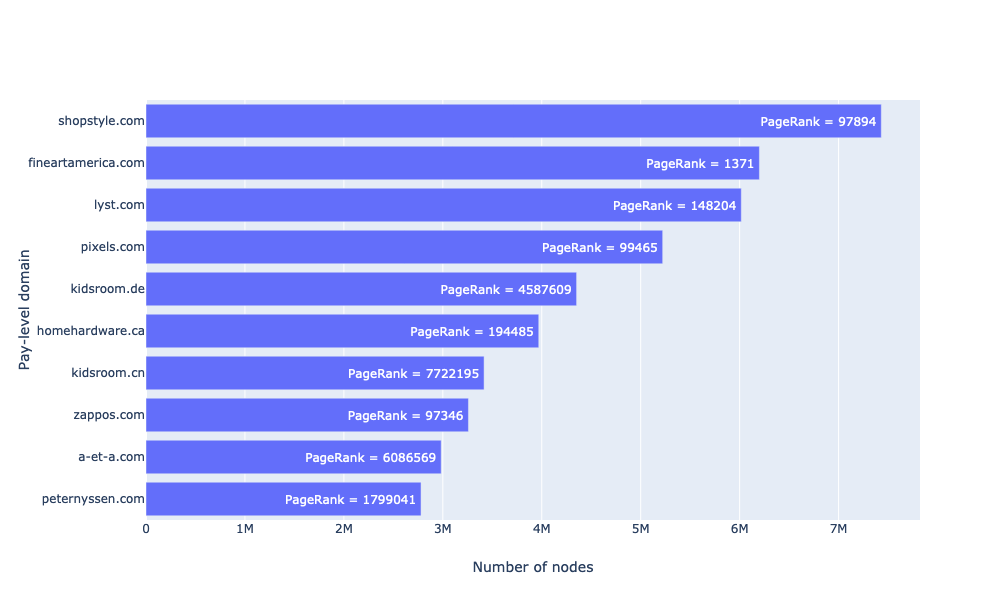}
  \caption{Count of the number of nodes associated with each pay-level domain along with Google's PageRank for that domain (only the 10 most frequent pay-level domains are shown) }
  \label{fig:domainranks}
  \Description{Count of the number of nodes associated with each pay-level domain along with Google's PageRank for that domain (only the 10 most frequent pay-level domains are shown) }
\end{figure*}

To study skewness at the level of pay-level domains, we attempted to determine if the most common websites have an impact on the type of product nodes that are present in the entire dataset. We found that the dataset contains nodes extracted from around 1 million pay-level domains and is not skewed towards having a significant portion of nodes associated with a particular pay-level domain. As can be seen from the ten common pay-level domains in Figure \ref{fig:domainranks}, node count associated with each of the ten pay-level domains obeys a roughly linear distribution. We further examined whether the nodes are crawled from trusted pay-level domains. A pay-level domain is defined by us as \emph{trusted domain} if it has low Google PageRank. Google PageRank judges the "value of a page" by measuring the quality and quantity of other pages that link to it. The main purpose of PageRank is to determine the relative importance or relative trust of a given page in the Web. So, a web page having a low PageRank is relatively more important or trustworthy than a web page having high PageRank. Hence, we computed the Google PageRanks\footnote{The Google PageRank was computed using the Open PageRank API, https://www.domcop.com/openpagerank/what-is-openpagerank} for the 10 common pay-level-domains with the results shown in Figure \ref{fig:domainranks}.

\subsection{Overall Data Completeness}\label{sec:datacompleteness}
When describing the raw data, we noted that even in compressed form, the entire Product dataset can be well over 100 GB. While this is easy to handle using a moderately sized Hadoop cluster, it is big enough that individual researchers and smaller organizations without access to such clusters (or operating on tight budgets) will be looking for approximate ways of extracting product nodes and their properties.

In fact, the observation that all the triples associated with a particular product node do not all appear \emph{together} in the Product dataset and may be spread out in the dataset incurs significant processing challenges in a non-parallel architecture.  

For certain downstream tasks, one may need all the triples associated with a particular node to be in memory. The easiest solution is to store the entire dataset in memory but this is not practical due to dataset size. To overcome this problem, we may need to sequentially read the triples associated with nodes from the dataset and have a window with a suitable size to ensure that we have `captured' all the triples associated with a particular node within that window.

More specifically, to find the extent of the `spread' of the triples associated with a particular node, we adopt the following mechanism. We attempt to discover if there is a \emph{window} of sufficiently small size within which we could find all the triples associated with a given node starting from the first triple associated with that  node, at least most of the time. The value of this size would  then tell us whether the dataset has sufficient \emph{localization} (i.e. less  spread in the sense defined above).

\begin{figure*}
  \centering
  \includegraphics[width=15cm, height=7cm]{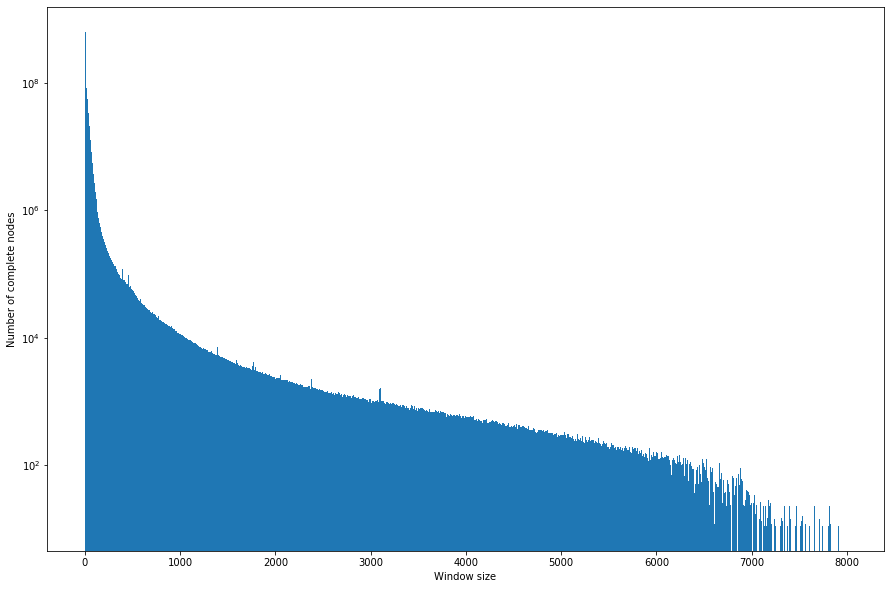}
  \caption{Number of complete nodes within the given window size. A complete node is one which has all triples associated with it occurring within a particular window size.}
  \label{fig:windowSizePlot}
  \Description{Number of complete nodes within a window of given size. A complete node is one which has all its associated triples occurring within a particular window size.}
\end{figure*}

To study localization in the Product dataset, we computed the minimum such window size for every product  node. The distribution of these window sizes associated with all the nodes in the Product dataset can be found in the Figure \ref{fig:windowSizePlot}. The average window size is found to be 27 and the 99th percentile value of window size distribution occurs at size 145. In other words, if we slide through the Product dataset with a window of size 145, starting from the first triple associated with a given node, we will be able to find all the triples associated with that node within the window containing the next 145 triples (at least for 99\% of the cases).


\section{Summary of Best Practices}\label{bestpractices}
This section succinctly lists some of the best practices determined through the previously described empirical studies for using the Product dataset in downstream applications and analyses. 

\begin{itemize}
    \item To have a broad coverage while extracting information from triples associated with specific property sets, we recommend considering variations and misspellings of concepts sets. Section \ref{sec:schemavar} shows that there are consistent sources of variations that could be identified. Other sources of variation, as well as more automated levels of detection, are left for future work. The use of powerful schema matching and entity resolution algorithms are clearly relevant here. 
    
    \item We recommend running simple heuristic  checks to confirm semantic validity of properties associated with the product nodes before ingesting property values (`objects') in downstream tasks. Some such rules were described earlier in Section \ref{sec:productchar} but many more opportunities for discovering other such rules (especially automatically) exist for future research. 
    
    \item If the downstream task is highly dependent on the language of the dataset, we recommend  not directly trusting the language tags associated with the text literals. From Section \ref{sec:dataskew}, we noted that using even a relatively straightforward language identifier model such as the one built using common fastText embeddings (to identify the language of a given text) may be more trustworthy than the explicit tag.
    
    \item For downstream tasks that need product nodes from different categories, we recommend selecting the subset of pay-level domains that cover all categories of products and extract products only from the triples associated with those pay-level domains. From Section \ref{sec:dataskew}, we can see that the distribution of pay-level domains is not skewed and downstream applications that use the data associated with such subsets have less likelihood of being biased.    
    
    \item If we need to extract trusted data from the Product dataset, we recommend limiting the extraction of information from triples associated with pay-level domains having a low PageRank. Low PageRank for a pay-level domain indicates that it is of high importance and trust. We could use a simple API to get PageRank for pay-level domains as in Section \ref{sec:dataskew}.
    
    \item To address concerns relating to Big Data, we recommend using a window of size 145 or more while sliding through the triples of the Product dataset. As shown in Section \ref{sec:datacompleteness}, having a window size of 145 ensures that we find all the properties associated with a given node within that window in 99\% of the cases.
\end{itemize}

\section{Conclusion and Future Work}\label{conclusion}
Schema.org and structured data have become highly significant in recent times. In this paper, we studied a product-specific schema.org dataset made recently available by the Web Data Commons project, and used a set of carefully designed empirical studies to devise a set of best practices that could be used by the research community to extract value from the raw data. We noted both minor concerns (such as variations in the manner in which schema.org concepts are represented in the Product dataset), as well as issues of semantic validity of properties and disagreement between explicitly stated language tags and the actual language of the underlying text literals. We recommended a set of best practices based on these findings. For example, we noted that, if the downstream task is highly dependent on the language of the dataset, directly trusting the tags is not the best course of action. Instead, one may want to use an ensemble based both on explicit tags and an automated language detection algorithm.

There are many avenues for future research, some of which we are actively pursuing. Certainly, the most promising avenue is to further refine our best practices and to identify potential subsets of the data where they may not be as applicable. We would also like to further study linguistic subsets of the data to assess whether data in certain languages are more prone to noise. Another aspect is temporality, namely, do our best practices also hold when consuming schema.org in other years? While the Common Crawl and WDC are generally very consistent in their crawling methodologies, significant changes have been incurred periodically. Finally, we would also like to study the graph-theoretic properties of the product-specific dataset, possibly using tools like Entity Resolution to complete the dataset. We would like to discover new best practices based on such analysis.

%
\bibliographystyle{ACM-Reference-Format}
\bibliography{data-quality-ref.bib}

\end{document}